\documentclass[vecphys]{svmult}
\usepackage{mathptmx}
\usepackage{amsmath}
\usepackage{amssymb}

\usepackage[bottom]{footmisc}

\usepackage{esint}

\makeindex

\begin{document}

\title*{Uncertainty related to position and momentum localization of a quantum state}

\author{{\L}ukasz Rudnicki}

\institute{Center for Theoretical Physics, Polish Academy of Sciences, Al. Lotnik{\'o}w 32/46\\
02-668 Warszawa, Poland\\
\email{rudnicki@cft.edu.pl}}

\maketitle

\abstract{This paper presents the uncertainty related to position and momentum localization of a quantum state in terms of entropic uncertainty relations. We slightly improve the inequality given in [Phys. Rev. A 74, 052101 (2006)] and introduce a new entropic measure with corresponding uncertainty relation.}

\section{Introduction}

From the famous Heisenberg uncertainty principle we can learn that ,,it is impossible to prepare states in which position
and momentum are simultaneously arbitrarily well localized'' \cite{rep}. It means that a measurement of two observables
dealing with position and momentum localization of a quantum state respectively shall expose some uncertainty. Our goal
is to describe this uncertainty (originating from complementarity of position and momentum variables) with the help of
the entropic uncertainty relations. To this end we shall use the R{\'e}nyi entropy defined for a set of probabilities
$P_{i}$:\begin{equation}
H_{\alpha}^{(P)}=\frac{1}{1-\alpha}\ln\left[\sum_{i}P_{i}^{\alpha}\right],\label{a1}\end{equation}
where $\alpha>0$. In some cases we will also use the Shannon entropy defined by the following limit:\begin{equation}
S^{(P)}=\lim_{\alpha\rightarrow1}H_{\alpha}^{(P)}=-\sum_{i}P_{i}\ln P_{i}.\label{a2}\end{equation}
To assure the maximal simplicity of the formulas appearing in this paper we will restrict ourselves to a one-dimensional
case, where the quantum state is described in the position representation by a normalized wave function $\psi(x)$. In the
momentum representation the same state is described by the wave function $\tilde{\psi}\left(p\right)$ related to the
previous one by the Fourier transformation:\begin{equation}
\tilde{\psi}\left(p\right)=\frac{1}{\sqrt{2\pi\hbar}}\int_{\mathbb{R}}dx\, e^{-ipx/\hbar}\psi\left(x\right).\label{b1}\end{equation}

To explain how to describe the uncertainty in terms of entropic uncertainty relations let us consider two arbitrary observables
($F$ and $G$) with corresponding hermitian operators $\hat{F}$ and $\hat{G}$. We assume that these operators have
point spectrum and in general do not commute with each other. The eigenstates $\varphi_{m}(x)$ of the operator $\hat{F}$, and $\theta_{n}(x)$
of the operator $\hat{G}$ form two orthonormal bases:\begin{equation}
\int_{\mathbb{R}}dx\,\varphi_{m}\left(x\right)\varphi_{m'}^{*}\left(x\right)=\delta_{mm'},\quad\int_{\mathbb{R}}dx\,\theta_{n}\left(x\right)\theta_{n'}^{*}\left(x\right)=\delta_{nn'}.\label{b2}\end{equation}
 The following completeness relations are also satisfied:\begin{equation}
\sum_{m}\varphi_{m}\left(x\right)\varphi_{m}^{*}\left(x'\right)=\sum_{n}\theta_{n}\left(x\right)\theta_{n}^{*}\left(x'\right)=\delta(x-x').\label{b3}\end{equation}
 With the observables $F$ and $G$ we can associate the probability distributions $\left|f_{m}\right|^{2}$ and $\left|g_{n}\right|^{2}$,
where:\begin{equation}
f_{m}=\int_{\mathbb{R}}dx\,\psi\left(x\right)\varphi_{m}^{*}\left(x\right),\label{b4}\end{equation}
 \begin{equation}
g_{n}=\int_{\mathbb{R}}dx\,\psi\left(x\right)\theta_{n}^{*}\left(x\right).\label{b5}\end{equation}
 The $f_{m}$ and $g_{n}$ vectors are connected by a unitary transformation $U_{mn}$:\begin{equation}
f_{m}=\sum_{n}U_{mn}\, g_{n},\quad g_{n}=\sum_{m}U_{mn}^{*}\, f_{m},\quad\sum_{n}U_{mn}U_{m'n}^{*}=\delta_{mm'},\label{b6}\end{equation}
 where:\begin{equation}
U_{mn}=\int_{\mathbb{R}}dx\,\varphi_{m}^{*}\left(x\right)\theta_{n}\left(x\right).\label{b7}\end{equation}
 According to Riesz theorem \cite{riesz,mu} we have the norm inequality:\begin{equation}
\left[c_{U}\sum_{m}\left|f_{m}\right|^{2\alpha}\right]^{1/\alpha}\leq\left[c_{U}\sum_{n}\left|g_{n}\right|^{2\beta}\right]^{1/\beta},\label{b8}\end{equation}
 \begin{equation}
\frac{1}{\alpha}+\frac{1}{\beta}=2,\quad\alpha\geq1,\quad c_{U}=\sup_{(m,n)}\left|U_{mn}\right|.\label{b9}\end{equation}
By definition $0\leq c_{U}\leq1$. Taking the logarithm of both sides of the norm inequality (\ref{b8}) we obtain the
uncertainty relation for the sum of the R{\'e}nyi entropies:\begin{equation}
H_{\alpha}^{(F)}+H_{\beta}^{(G)}\geq-2\ln c_{U}.\label{b10}\end{equation}
For every given pair of observables $(F,G)$ one can find the eigenstates $\varphi_{m}$ and $\theta_{n}$ and calculate the value
of $c_{U}$. Typically, when the commutator of $\hat{F}$ and $\hat{G}$ does not vanish, $c_{U}$ shall be less than
$1$, what simply means that the values of the observables $F$ and $G$ cannot be simultaneously well determined. When this
commutator is equal to $0$ the operators $\hat{F}$ and $\hat{G}$ share common eigenvectors and $c_{U}=1$.

\section{Localized measurements}

To include information about localization properties of the quantum state we shall change our previous set of observables.
Let us divide both position and momentum space into equal bins. To describe this partition we introduce the characteristic
function:\begin{equation}
\chi_{j}\left(s\right)=\begin{cases}
1 & s\in\left[\left(j-\frac{1}{2}\right)\delta s,\,\left(j+\frac{1}{2}\right)\delta s\right]\\
0 & \textrm{elsewhere}\end{cases}.\label{c1}\end{equation}
The index $j$ labels the bins and $\delta s$ denotes the bin's width. With the $k$th bin in the position space we associate
an observable $A_{k}$ with corresponding eigenstates $\varphi_{km}(x)$ forming in the line segment, described by $\chi_{k}\left(x\right)$,
the orthonormal basis:\begin{equation}
\int_{\mathbb{R}}dx\,\chi_{k}\left(x\right)\varphi_{km}\left(x\right)\varphi_{km'}^{*}\left(x\right)=\delta_{mm'},\quad\sum_{m}\varphi_{km}\left(x\right)\varphi_{km}^{*}\left(x'\right)=\delta_{k}(x-x').\label{c2}\end{equation}
One should notice that the Dirac delta function in (\ref{b3}) differs from the one in (\ref{c2}) by its domain. The
domain of $\delta_{k}(x-x')$ is restricted to the $k$th bin. Similarly in the $l$th bin in the momentum space, described
by $\chi_{l}\left(p\right)$, we introduce an observable $B_{l}$ with eigenstates $\theta_{ln}(p)$. We have:\begin{equation}
\int_{\mathbb{R}}dp\,\chi_{l}\left(p\right)\theta_{ln}\left(p\right)\theta_{ln'}^{*}\left(p\right)=\delta_{nn'},\quad\sum_{n}\theta_{ln}\left(p\right)\theta_{ln}^{*}\left(p'\right)=\delta_{l}(p-p').\label{c3}\end{equation}
For the observables $A_{k}$ and $B_{l}$ we have the probability distributions $\left|a_{km}\right|^{2}$ and $\left|b_{ln}\right|^{2}$,
where:\begin{equation}
a_{km}=\int_{\mathbb{R}}dx\,\chi_{k}\left(x\right)\psi\left(x\right)\varphi_{km}^{*}\left(x\right),\label{c4}\end{equation}
 \begin{equation}
b_{ln}=\int_{\mathbb{R}}dp\,\chi_{l}\left(p\right)\tilde{\psi}\left(p\right)\theta_{ln}^{*}\left(p\right).\label{c5}\end{equation}
In this case the unitary transformation $U_{kmln}$:\begin{equation}
a_{km}=\sum_{l,n}U_{kmln}\, b_{ln},\quad b_{ln}=\sum_{k,m}U_{kmln}^{*}\, a_{km},\label{c6}\end{equation}
reads:\begin{equation}
U_{kmln}=\frac{1}{\sqrt{2\pi\hbar}}\int_{\mathbb{R}}dx\,\chi_{k}\left(x\right)\int_{\mathbb{R}}dp\,\chi_{l}\left(p\right)e^{ipx/\hbar}\varphi_{km}^{*}\left(x\right)\theta_{ln}\left(p\right).\label{c7}\end{equation}
Using the same arguments as before we obtain the uncertainty relation: \begin{equation}
H_{\alpha}^{(A)}+H_{\beta}^{(B)}\geq-2\ln c_{U},\quad c_{U}=\sup_{(k,m,l,n)}\left|U_{kmln}\right|,\label{c8}\end{equation}
where now the R{\'e}nyi entropies are:\begin{equation}
H_{\alpha}^{(A)}=\frac{1}{1-\alpha}\ln\left[\sum_{k,m=-\infty}^{\infty}\left|a_{km}\right|^{2\alpha}\right],\quad H_{\beta}^{(B)}=\frac{1}{1-\beta}\ln\left[\sum_{l,n=-\infty}^{\infty}\left|b_{ln}\right|^{2\beta}\right].\label{c9}\end{equation}
The uncertainty relation (\ref{c8}) caries both information about the uncertainty of position and momentum localization,
and uncertainties associated with the observables $A_{k}$ and $B_{l}$. To calculate the uncertainty related only to
the localization properties of the state we shall find the maximal value $C_{max}$ among $c_{U}$'s calculated for whole possible choices of $\varphi_{km}\left(x\right)$
and $\theta_{ln}\left(p\right)$. To do this we will firstly use the H{\"o}lder inequality \cite{mit}:\begin{align}
\left|U_{kmln}\right| & \leq\frac{1}{\sqrt{2\pi\hbar}}\left(\int_{\mathbb{R}}dx\,\chi_{k}\left(x\right)\left|\varphi_{km}\left(x\right)\right|^{2}\right)^{1/2}\times\label{c10}\\
 & \left(\int_{\mathbb{R}}dx\,\chi_{k}\left(x\right)\left|\int_{\mathbb{R}}dp\,\chi_{l}\left(p\right)e^{ipx/\hbar}\theta_{ln}\left(p\right)\right|^{2}\right)^{1/2}.\nonumber \end{align}
From the orthonormality condition (\ref{c2}) we obtain that the integral in the first parenthesis is equal to $1$. Now
we rewrite (\ref{c10}):\begin{equation}
\left|U_{kmln}\right|\leq\sqrt{W\left[\theta_{ln}\right]},\label{c11}\end{equation}
where we have introduced the functional $W\left[\theta_{ln}\right]$ of the form:\begin{equation}
W\left[\theta_{ln}\right]=\int_{\mathbb{R}}dp\,\chi_{l}\left(p\right)\int_{\mathbb{R}}dq\,\chi_{l}\left(q\right)Q_{k}(p-q)\theta_{ln}\left(p\right)\theta_{ln}^{*}\left(q\right),\label{c12}\end{equation}
and the integral kernel $Q_{k}(p-q)$ is: \begin{equation}
Q_{k}(p-q)=\frac{1}{2\pi\hbar}\int_{\mathbb{R}}dx\,\chi_{k}\left(x\right)e^{i(p-q)x/\hbar}=e^{-i\frac{k\delta x}{\hbar}(p-q)}\frac{\sin\left[\frac{\delta x}{2\hbar}\left(p-q\right)\right]}{\pi(p-q)}.\label{c13}\end{equation}
We can define a new function:\begin{equation}
\Psi(p)=e^{-i\frac{k\delta x}{\hbar}\left(p+l\delta p\right)}\theta_{ln}\left(p+l\delta p\right),\label{c14}\end{equation}
in terms of which the functional $W$ takes a simpler form:\begin{equation}
W\left[\Psi\right]=\int_{\mathbb{R}}dp\,\chi_{0}\left(p\right)\int_{\mathbb{R}}dq\,\chi_{0}\left(q\right)Q_{0}(p-q)\Psi\left(p\right)\Psi^{*}\left(q\right).\label{c15}\end{equation}
We have got rid of the indices $k$ and $l$ what could be done due to the translational symmetries both in coordinate
and momentum spaces. To find the maximal possible value of the functional $W\left[\Psi\right]$ we have to solve the variational
equation:\begin{equation}
\frac{\delta}{\delta\Psi}\left(W\left[\Psi\right]-\lambda\int_{\mathbb{R}}dp\,\chi_{0}\left(p\right)\left|\Psi\left(p\right)\right|^{2}\right)=0,\label{c16}\end{equation}
where $\lambda$ is the Lagrange multiplier associated with the normalization constraint. The equation (\ref{c16}) leads
to the Fredholm integral equation of the second kind \cite{schur}:\begin{equation}
\frac{1}{\pi}\int_{-1}^{1}ds\,\frac{\sin\left[\frac{\gamma}{4}\left(t-s\right)\right]}{t-s}\Psi_{j}\left(s\right)=\lambda_{j}\Psi_{j}\left(t\right),\label{c17}\end{equation}
where $\gamma=\delta x\delta p/\hbar$ is a dimensionless parameter which appears instead of the bin's widths $\delta x$
and $\delta p$. The maximal value of $W\left[\Psi\right]$ is the largest eigenvalue $\lambda_{0}$ of (\ref{c17}) \cite{rep,schur}:\begin{equation}
W\left[\Psi\right]\leq\lambda_{0}=\frac{\gamma}{2\pi}\left[R_{00}\left(\frac{\gamma}{4},1\right)\right]^{2},\label{c18}\end{equation}
where $R_{00}\left(s,t\right)$ is one of the radial prolate spheroidal wave functions of the first kind%
\footnote{in the Wolfram Mathematica's notation it is $\mathtt{SpheroidalS1[0,0,s,t]}$.%
} \cite{abr}. Thus, with the help of the variational method we have found that:\begin{equation}
C_{max}=\sqrt{\lambda_{0}}=\sqrt{\frac{\gamma}{2\pi}}R_{00}\left(\frac{\gamma}{4},1\right).\label{c19}\end{equation}
For every finite value of the parameter $\gamma$ we have $C_{max}<1$ which is the measure of uncertainty of position
and momentum localization. The corresponding entropic uncertainty relation is the Maassen-Uffink type inequality \cite{mu,sen}:\begin{equation}
H_{\alpha}^{(A)}+H_{\beta}^{(B)}\geq-2\ln C_{max}.\label{c20}\end{equation}

\section{Probability distributions for localization in coordinate and momentum space}

The probability distributions $\left|a_{km}\right|^{2}$ and $\left|b_{ln}\right|^{2}$ carry both information about localization
and information about the observables $A_{k}$ and $B_{l}$. In order to obtain a pure information about localization
we shall trace out the observables' degrees of freedom:\begin{equation}
q_{k}=\sum_{m}\left|a_{km}\right|^{2}=\int_{\mathbb{R}}dx\,\chi_{k}\left(x\right)\left|\psi\left(x\right)\right|^{2},\label{c21}\end{equation}
 \begin{equation}
p_{l}=\sum_{n}\left|b_{ln}\right|^{2}=\int_{\mathbb{R}}dp\,\chi_{l}\left(p\right)\left|\tilde{\psi}\left(p\right)\right|^{2}.\label{c22}\end{equation}
The $q_{k}$ coefficient has an interpretation of the probability of finding the particle in the $k$'th bin in the coordinate
space. The $p_{l}$ coefficient can be interpreted in a similar manner. The probability distributions $q_{k}$ and $p_{l}$
are affected by the Heisenberg uncertainty principle, thus we have \cite{rep}:\begin{equation}
\vee_{k,l}\; q_{k}+p_{l}\leq1+\sqrt{\lambda_{0}},\label{c23}\end{equation}
where $\lambda_{0}$ was defined in (\ref{c18}). The way of finding the relation (\ref{c23}) leads to the same integral
equation (\ref{c17}). From (\ref{c23}) we can easily find that:\begin{equation}
\vee_{k,l}\; q_{k}p_{l}\leq\frac{1}{4}\left(1+\sqrt{\lambda_{0}}\right)^{2}.\label{c24}\end{equation}
The inequality (\ref{c24}) is suitable to be used together with the Shannon entropies, because the sum of the Shannon
entropies is:\begin{equation}
S^{(q)}+S^{(p)}=-\sum_{k,l}q_{k}p_{l}\ln q_{k}p_{l},\label{c25}\end{equation}
and we are able to find a simple lower bound:\begin{equation}
S^{(q)}+S^{(p)}\geq-\sum_{k,l}q_{k}p_{l}\ln\left[\frac{1}{4}\left(1+\sqrt{\lambda_{0}}\right)^{2}\right]\geq-2\ln\left[\frac{1}{2}\left(1+C_{max}\right)\right].\label{c26}\end{equation}
We can call it the Deutsch type inequality \cite{sen,d}.

In spite of the fact that $H_{\alpha}^{(A)}\geq H_{\alpha}^{(q)}$ and $H_{\beta}^{(B)}\geq H_{\beta}^{(p)}$ there is
another lower bound for the sum $H_{\alpha}^{(q)}+H_{\beta}^{(p)}$ which in the quantum regime ($\gamma<1$) is significantly
better than (\ref{c20}). The probability distributions (\ref{c21}, \ref{c22}) fulfill the chain of inequalities:\begin{eqnarray}
\left(\sum_{k=-\infty}^{\infty}q_{k}^{\alpha}\right)^{1/\alpha} & \leq & \left(\delta x\right)^{1-1/\alpha}\left(\int_{\mathbb{R}}dx\left|\psi(x)\right|^{2\alpha}\right)^{1/\alpha}\nonumber \\
 & \leq & \left(\delta x\right)^{(1-1/\alpha)}\left(\frac{\alpha}{\pi}\right)^{-1/2\alpha}\left(\frac{\beta}{\pi}\right)^{1/2\beta}\left(\int_{\mathbb{R}}dp\left|\tilde{\psi}(p)\right|^{2\beta}\right)^{1/\beta}\label{c27}\\
 & \leq & \left(\delta x\right)^{(1-1/\alpha)}\left(\delta p\right)^{(1/\beta-1)}\left(\frac{\alpha}{\pi}\right)^{-1/2\alpha}\left(\frac{\beta}{\pi}\right)^{1/2\beta}\left(\sum_{l=-\infty}^{\infty}p_{l}^{\beta}\right)^{1/\beta},\nonumber \end{eqnarray}
where the first and the last one are the Jensen inequalities and in the middle there is the famous Beckner inequality
\cite{beck}. From (\ref{c27}) one can derive the uncertainty relation \cite{ibbur}:\begin{equation}
H_{\alpha}^{(q)}+H_{\beta}^{(p)}\geq-\frac{1}{2}\left(\frac{\ln\alpha}{1-\alpha}+\frac{\ln\beta}{1-\beta}\right)+\ln\pi-\ln\gamma.\label{c28}\end{equation}
This lower bound is better than (\ref{c20}) and (\ref{c26}) for small values of $\gamma$ parameter but unfortunately
it becomes negative (and in fact meaningless) for:\begin{equation}
\gamma\geq\frac{\pi}{\alpha}\left(2\alpha-1\right)^{\frac{2\alpha-1}{2(\alpha-1)}}.\label{c29}\end{equation}
In the case of the Shannon entropies ($\alpha\rightarrow1)$ the condition (\ref{c29}) reads $\gamma\geq e\pi$.

\section{Summary}

The best calculated lower bounds for the sum of Shannon entropies which can be treated as a measure of uncertainty of
position and momentum localization of the quantum state are:\begin{equation}
S^{(A)}+S^{(B)}\geq-\ln\left[\min\left\{ \frac{\gamma}{e\pi},\frac{\gamma}{2\pi}\left[R_{00}\left(\frac{\gamma}{4},1\right)\right]^{2}\right\} \right],\end{equation}
in the case of the probability distributions $\left|a_{km}\right|^{2}$ and $\left|b_{ln}\right|^{2}$, and:\begin{equation}
S^{(q)}+S^{(p)}\geq-\ln\left[\min\left\{ \frac{\gamma}{e\pi},\frac{1}{4}\left[1+\sqrt{\frac{\gamma}{2\pi}}R_{00}\left(\frac{\gamma}{4},1\right)\right]^{2}\right\} \right],\end{equation}
in the case of the probability distributions (\ref{c21}, \ref{c22}) which completely describe the localization properties.

\begin{acknowledgement}
I would like to thank Iwo Bialynicki-Birula who was the supervisor of this work. This research was supported by the grant from the Polish Ministry of Science and Higher Education.
\end{acknowledgement}

\end{document}